\documentclass[12pt]{article}
\begin{document}
\begin{flushright}
{hep-th/0401104}
\end{flushright}

\begin{center}
{\large {\bf Mecanismo de autointeracci\'on en \\ el modelo masivo
vectorial
de Hagen}} \\
\vskip 12pt {\bf P\'{\i}o J. Arias $^{a,b}$
\footnote{email:parias@fisica.ciens.ucv.ve}}\\
$^{a}${\it Centro de F\'{\i}sica Te\'orica y Computacional,
Facultad de
Ciencias, U.C.V., AP 47270, Caracas 1041-A, Venezuela} \\
$^{b}${\it Centro de Astrof\'{\i}sica Te\'orica, Facultad de
Ciencias, U.L.A., La Hechicera, M\'erida 5101, Venezuela}\\
\end{center}

\begin{abstract}
It is shown that the non-abelian vectorial model, proposed by
C.R.Hagen is obtained using the self-interaction mechanism. The
equivalence between this model and the non-abelian topologically
massive one is studied showing that the existing equivalence in
the abelian models is not sustained.
\end{abstract}


\vskip 1truecm
\begin{narrower}
Se muestra que el modelo masivo vectorial no-abeliano, propuesto
por C.R.Hagen, se obtiene usando el mecanismo de
autointeracci\'on. Se estudia la equivalencia de este modelo con
el modelo topol\'ogico masivo no-abeliano y se obtiene que la
equivalencia existente a nivel abeliano no se mantiene.
\end{narrower}
\vskip 0.5truecm
 Palabras clave:{\it Mecanismo de autointeracci\'on,
Modelo no-abelianos} \vskip 12pt {\bf PACS: $11.10.Ef$,$11.10.Kk$,
$11.15.Kc$, $11.30.Fs$} \vskip 0.5truecm

La teor\'{\i}a de campos en 2+1 dimensiones constituye un
excelente escenario para el entendimiento y estudio de
teor\'{\i}as f\'{\i}sicas en dimensi\'on 3+1. \'Esto por su
sencillez y adem\'as por haber provisto nuevas ideas al estudio de
la f\'{\i}sica en 3+1 dimensiones. Es as\'{\i} como los modelos en
2+1 dimensiones han estado motivados por sus posibles aplicaciones
en el efecto Hall fraccionario, la superconductividad a altas
temperaturas y los procesos en presencia de cuerdas c\'osmicas.

La estructura de las teor\'{\i}as no-abelianas puede obtenerse
f\'{\i}sicamente bajo el requerimiento de que se acople
consistentemente a fuentes din\'amicas siguiendo el mecanismo de
autointeracci\'on. Para esto comenzamos con una teor\'{\i}a
vectorial o tensorial la cual posee alguna invarancia de calibre
que asegure la no propagaci\'on de los campos asociados a los
spines menores al que se quiere describir. La identidad de Biachi
asociada a esta invariancia requiere que las fuentes del campo
sean conservadas. \'Esta conservaci\'on se pierde al acoplar
din\'amicamente al campo, a menos que est\'e autoacoplado. El auto
acoplamineto requerido es determinado a partir de la corriente de
N\"oether asociada con la invariancia global interna presente en
estos modelos. Aplicando este mecanismo se obtiene de forma
natural la acciones de Yang-Mills, de Einstein y de
supeergravedad\cite{1,2}, as\'{\i} como las acciones de la
teor\'{\i}a topol\'ogica masiva no-abeliana\cite{3}, la de
Chapline-Manton\cite{4}, la de Freedman-Townsend\cite{5} y la del
modelo masivo autodual no-abeliano en 2+1 dimensiones\cite{6},
entre otros.

En este trabajo mostraremos como el modelo de Hagen para una
part\'{\i}cula de spin 1 masivo en 2+1 dimensiones\cite{7} est\'a
conectado con su versi\'on no-abeliana\cite{8} usando el mecanismo
de autointeracci\'on antes expuesto. El modelo a considerar es
equivalente a los modelos autodual y topol\'ogico masivo a nivel
abeliano\cite{9}. Sin ambargo, resulta ser no equivalente a nivel
no-abeliano dado que los t\'erminos de autointeracci\'on
proporcionanan correcciones adicionales en el analisis cu\'antico
perturbativo\cite{8}.

En 2+1 dimensiones existen distintas descripciones para una
part\'{\i}cula masiva con spin 1. El modelo mas sencillo es el
modelo autodual\cite{10} descrito por la acci\'on
\begin{equation}
  S_{AD} = - \frac{m}{2}\int \,d^{3}x \,\left(\varepsilon^{{\mu}{\nu}{\lambda}}a_{\mu}\partial_{\nu}a_{\lambda}
+ ma_{\mu}a^{\mu}\right),
\end{equation}
donde $\varepsilon^{012}=1$ y $\eta_{\mu\nu}=(-++)$. \'Este modelo
no posee invariancias locales y puede mostrarse que constituye una
versi\'on del modelo topol\'ogico masivo luego de fijar
convenientemente el calibre\cite{11,12,13,ban}. El modelo
topol\'ogico masivo viene descrito por la acci\'on\cite{14}
\begin{equation}
  S_{TM} = - \frac{1}{2}\int \,d^{3}x \,\left(\frac{1}{2}F_{\mu\nu}F^{\mu\nu}-
  m\varepsilon^{{\mu}{\nu}{\lambda}}a_{\mu}\partial_{\nu}a_{\lambda}\right),
\end{equation}
con $F_{\mu\nu}=\partial_{\mu}a_{\nu}-\partial_{\nu}a_{\mu}$,
donde el primer t\'ermino corresponde al conocido t\'ermino de
Maxwell y el segundo se conoce como el t\'ermino de Chern-Simons
vectorial. Las ecuaciones de movimiento de este modelo poseen la
misma invariancia que la electrodin\'amica usual. Este hecho
resulta interesante pues presenta el ejemplo de una teor\'{\i}a
masiva que posee invariancia de calibre. Tal como apuntamos
anteriormente los modelos autodual y topol\'ogico masivo estan
conectados por una fijaci\'on de calibre. Sin embargo, puede verse
que los espacios de soluciones difieren en soluciones de
car\'acter topol\'ogico y que estan conectados por una
transformaci\'on de dualidad\cite{15, 16, 17, wot1, 18, wot2}.

Otra descripci\'on para una part\'{\i}cula masiva con spin 1 la
proporciona la acci\'on propuesta por C.R.Hagen\cite{7}
\begin{equation}
\label{Sh} S_{H}=\frac{1}{2}\int d^{3}x\Bigl(-f^{\mu}f_{\mu}+
  2\varepsilon^{{\mu}{\nu}{\lambda}}f_{\mu}\partial_{\nu}a_{\lambda}
+m\varepsilon^{{\mu}{\nu}{\lambda}}a_{\mu}\partial_{\nu}a_{\lambda}
  +\frac{\lambda}{m}\varepsilon^{{\mu}{\nu}{\lambda}}f_{\mu}\partial_{\nu}f_{\lambda}\Bigr).
\end{equation}
Esta acci\'on se convierte en la de la topol\'ogima masiva, a
primer orden, si tomamos $\lambda=0$. Adem\'as puede mostrarse que
cuando $\lambda=1$ no tiene din\'amica local. La masa de las
excitaciones es $\vert m/(1-\lambda)\vert$. Para ver esto y su
equivalencia con los modelos autodual y topol\'ogico masivo
analicemos las ecuaciones de movimiento de (\ref{Sh})
\begin{eqnarray}
-f^{\mu}+\varepsilon^{\mu\nu\lambda}\partial_{\nu}\bigl(a_{\lambda}+\frac{\lambda}{m}f_{\lambda}\bigr)=0,\\
\varepsilon^{\mu\nu\lambda}\partial_{\nu}\bigl(f_{\lambda}+ma_{\lambda}\bigr)=0.
\end{eqnarray}
En este sistema es claro que si $\lambda=0$ el sitema se
transforma en el de la teor\'{\i}a topol\'ogica masiva. Por otro
lado en la segunda de estas ecuaciones observamos que localmente
$f_{\lambda}+ma_{\lambda}=\partial_{\lambda}\rho$, si fijamos
calibre de tal forma que $f_{\lambda}+ma_{\lambda}=0$ y vamos a la
primera ecuaci\'on el sistema corresponder\'{\i}a al de la
teor\'{\i}a autodual con masa $m/(1-\lambda)$. Si $\lambda=1$ el
sistema s\'olo describe estados globales que corresponden a
$F_{\mu\nu}=0$.

Para completar de analizar la cinem\'atica de $S_{H}$ pasamos a
obtener la acci\'on reducida. Para esto tomamos
\begin{equation}
\Phi_0=\Phi \quad,\quad
\Phi_{i}=\varepsilon_{{i}{j}}\partial_{j}\Phi^{T}+\partial_{i}a\Phi^{L},
\end{equation}
donde $\Phi_{\mu}\equiv(a_{\mu},f_{\mu})$, $i,j=1,2$ y
$\varepsilon_{12}=1$. Al sustituir esta descomposici\'on en
$S_{H}$ observaremos que $a$ y $f$ son multiplicadores de Lagrange
asociados a los v\'{\i}nculos $f^T=ma^T$ y $f=(1-\lambda)\Delta
a^T$. Teniendo esto en cuenta llegamos a
\begin{eqnarray}
\label{red0} S_{H}&=& \frac{1}{2}\int d^{3}x
\Bigl((1-\lambda)\dot{a}^{T}(-\Delta)f^{L}+f^{L}(-\Delta)f^{L}\nonumber \\
&&\qquad \qquad
-{(1-\lambda)}^2(-\Delta)a^{T}(-\Delta)a^{T}-m^{2}a^{T}(-\Delta)a^{T}
\Bigr),
\end{eqnarray}
con $\Delta=\partial_i\partial_i$. A este nivel notamos que si
$\lambda=1$ no hay propagaci\'on alguna de los campos, adem\'as si
sustituimos $(1-\lambda)a^T\to{a^T}$ y
$\frac{m}{(1-\lambda)}\to{m}$ la acci\'on corresponder\'{\i}a a la
que hubi\'esemos obtenido si partieramos de la acci\'on
topol\'ogica masiva.

Si sustituimos $Q=(1-\lambda){(-\Delta)}^{1/2}a^T$,
$\Pi={(-\Delta)}^{1/2}f^L$ en (\ref{red0}) llegamos a la acci\'on
reducida
\begin{equation}
S_{H}^{red}=\int d^{3}x\Bigl[[\Pi\dot{Q}-\frac{1}{2}\Pi\,\Pi
-\frac{1}{2}Q\left(-\Delta+\bigl(\frac{m}{(1-\lambda)}\bigr)^2\right)Q\Bigr],
\end{equation}
donde queda claro que la teor\'{\i}a describe una excitaci\'on de
masa $\vert m/(1-\lambda)\vert$ con energ\'{\i}a definida
positiva.

Pasamos ahora a aplicar el mecanismo de autointeracci\'on a partir
de $S_H$ en (\ref{Sh})\cite{1,2}. La conexi\'on con el modelo
topol\'ogico masivo se obtiene tomando $ \lambda=0$. Partimos con
un conjunto de campos $a_{\mu}^a$ y $f_{\mu}^a$, con $a=1,2,3$,
pensando en una invariancia bajo rotaciones r\'{\i}gidas (a la
SU(2)). Este proceder permite adoptar una notaci\'on vectorial que
resultar\'a mas simple. La generalizaci\'on a otro tipo de grupos
de invariancia se realizar\'{\i}a de manera an\'aloga. La acci\'on
de partida es
\begin{equation}
\label{Sh0}
S_{0}=\frac{1}{2}\int d^{3}x
\Bigl(-\vec{f}^{\mu}\cdot\vec{f}_{\mu}+
  2\varepsilon^{{\mu}{\nu}{\lambda}}\vec{f}_{\mu}\cdot\partial_{\nu}\vec{a}_{\lambda}
  + m\varepsilon^{{\mu}{\nu}{\lambda}}\vec{a}_{\mu}\cdot\partial_{\nu}\vec{a}_{\lambda}
  +\frac{\lambda}{m}\varepsilon^{{\mu}{\nu}{\lambda}}\vec{f}_{\mu}\cdot\partial_{\nu}\vec{f}_{\lambda}\Bigr),
\end{equation}
la cual es invariante bajo los cambios globales
\begin{equation}
\label{global}
\vec{\Phi}_{\mu}\to
\vec{\Phi}_{\mu}+\vec{\omega}\times\vec{\Phi}_{\mu},
\end{equation}
y sus ecuaciones de movimiento poseen la invariancia de calibre
\begin{equation}
\delta\vec{a}_{\mu}=\partial_{\mu}\vec{\Lambda}(x)\qquad,
\qquad\delta\vec{f}^{\mu}=0.
\end{equation}

La corriente de N\"oether asociada a la inavariancia global es
\begin{equation}
\vec{j}^{\mu}=\varepsilon^{\mu\nu\lambda}\left[(\vec{f}_{\nu}+
\frac{m}{2}\vec{a}_{\nu})\times\vec{a}_{\lambda}+\frac{\lambda}{2m}\vec{f}_{\nu}\times\vec{f}_{\lambda}\right],
\end{equation}
la cual se conserva si usamos las ecuaciones de movimiento que
surgen de (\ref{Sh0}).

Ahora sumamos a $S_0$ un t\'ermino de autointeracci\'on de forma
tal que al hacer variaciones en \'este respecto a $\vec{a}_{\mu}$
se reobtenga $\vec{j}^{\mu}$. Este t\'ermino de autointeracci\'on
es
\begin{equation}
S^{int}=-\frac{g}{2}\int
d^3x\Bigl[\varepsilon^{\mu\nu\lambda}\Bigl(\vec{a}_{\mu}\cdot\bigl(\vec{f}_{\nu}+
\frac{m}{3}\vec{a}_{\nu}\bigr)\times\vec{a}_{\lambda}
+\frac{\lambda}{m}\vec{a}_{\mu}\cdot\vec{f}_{\nu}\times\vec{f}_{\lambda}\Bigr)+
F[f_{\mu}]\Bigr] ,
\end{equation}
donde $g$ es un par\'ametro de acoplamiento con unidades
$L^{-1/2}$ y adem\'as hemos indicado la posibilidad de adicionar
t\'erminos que dependan solamente de las $f$´s, ya que lo que
requerimos es que $\delta
S^{int}/\delta\vec{a}_{\mu}=\vec{j}^{\mu}$. Un t\'ermino posible,
no cuadr\'atico, covariante, invariante de calibre y bajo
(\ref{global}) ser\'{\i}a de la forma
$\sim\varepsilon^{\mu\nu\lambda}\vec{f}_{\mu}\cdot\vec{f}_{\nu}\times\vec{f}_{\lambda}$.
As\'{\i} la acci\'on que resulta de sumar a $S_0$ el t\'ermino de
autointeracci\'on, con el t\'ermino de las $f$´s propuesto,
tendr\'a la forma
\begin{eqnarray}
\label{S1} &&S=\frac{1}{2}\int
d^3x\Bigl[-\vec{f}^{\mu}\cdot\vec{f}_{\mu}+
\varepsilon^{\mu\nu\lambda}\vec{f}_{\mu}\cdot\bigl(\partial_{\nu}\vec{a}_{\lambda}
-\partial_{\lambda}\vec{a}_{\nu}-g\vec{a}_{\nu}\times\vec{a}_{\lambda}\bigr)\nonumber\\
&&\qquad\qquad+m\varepsilon^{\mu\nu\lambda}\bigl(\vec{a}_{\mu}\partial_{\nu}\vec{a}_{\lambda}-
\frac{g}{3}\vec{a}_{\mu}\cdot\vec{a}_{\nu}\times\vec{a}_{\lambda}\bigr)\nonumber\\
&&\qquad\qquad+\frac{\lambda}{m}\varepsilon^{\mu\nu\lambda}\vec{f}_{\mu}\cdot\bigl(\partial_{\nu}\vec{f}_{\lambda}
-g\vec{a}_{\nu}\times\vec{f}_{\lambda}\bigr)-\frac{g\kappa}{3m^2}
\varepsilon^{\mu\nu\lambda}\vec{f}_{\mu}\cdot\vec{f}_{\nu}\times\vec{f}_{\lambda}\Bigr],
\end{eqnarray}
donde $\kappa$ es un constante adimensionada. En (\ref{S1}) los
tres primeros t\'erminos corresponden a la acci\'on topol\'ogica
masiva no-abeliana. Los dos \'ultimos t\'erminos
corresponder\'{\i}an a la contribuci\'on adicional en la
teor\'{\i}a de Hagen no-abeliana tal como la propuso en la
referencia\cite{8} . Para establecer conexi\'on con la
formulaci\'on usual de teor\'{\i}as no-abelianas, pensamos en
generadores $T^a$ antiherm\'{\i}ticos que satisfacen
$trT^aT^b=-\frac{1}{2}\delta^{ab}$, los cuales act\'uan en la
representaci\'on adjunta del grupo. Para estos
$[T^a,T^b]=f^{abc}T^c$, donde $f^{abc}$ son las constantes de
estructura del grupo (es el caso de SU(2) \'estas son
$\varepsilon^{abc})$) y los campos los representamos
matricialmente como $\Phi_{\mu}=gT^a\Phi_{\mu}^a$. Con esta
notaci\'on (\ref{S1}) se escribe de forma compacta como
\begin{eqnarray}
\label{Snab} S_{H}^{na}=\frac{1}{g^2}\int d^3x\,
tr\Bigl[{f}^{\mu}{f}_{\mu}-\varepsilon^{\mu\nu\lambda}{f}_{\mu}F_{\nu\lambda}(a)
-m\varepsilon^{\mu\nu\lambda}\bigl({a}_{\mu}\partial_{\nu}{a}_{\lambda}-
\frac{2}{3}{a}_{\mu}{a}_{\nu}{a}_{\lambda}\bigr)\nonumber\\
-\frac{\lambda}{m}\varepsilon^{\mu\nu\lambda}{f}_{\mu}{\cal{D}}_{\nu}{f}_{\lambda}
+\frac{2\kappa}{3{m}^2}\varepsilon^{\mu\nu\lambda}{f}_{\mu}{f}_{\nu}{f}_{\lambda}\Bigr],
\end{eqnarray}
donde $F_{\nu\lambda}(a)=\partial_{\nu}{a}_{\lambda}
-\partial_{\lambda}{a}_{\nu}-[{a}_{\nu},{a}_{\lambda}]$ y
${\cal{D}}_{\nu}f_{\lambda}=\partial_{\nu}f_{\lambda}-[a_{\nu},f_{\lambda}]$
es la derivada covariante de $f_{\lambda}$ bajo las
transformaciones de calibre
\begin{equation}
\delta a_{\mu}={\cal{D}}_{\mu}\omega(x) \qquad,\qquad \delta
f^{\mu}=[\omega(x),f^{\mu}],
\end{equation}
con $\omega(x)=gT^a\omega^a(x)$, las cuales dejan invariantes las
ecuaciones de movimiento de (\ref{Snab}). Estas \'ultimas, en caso
de existir alg\'un acoplamiento externo, resultan ser
\begin{eqnarray}
\label{em}
f^{\mu}-{}^{\star}f^{\mu}(a)-
\frac{\lambda}{m}\varepsilon^{\mu\nu\lambda}{\cal{D}}_{\nu}{f}_{\lambda}
+\frac{\kappa}{2m^2}\varepsilon^{\mu\nu\lambda}[{f}_{\nu},{f}_{\lambda}]&=&0,\nonumber \\
m{}^{\star}f^{\mu}(a)+
\varepsilon^{\mu\nu\lambda}{\cal{D}}_{\nu}{f}_{\lambda}
-\frac{\lambda}{2m}\varepsilon^{\mu\nu\lambda}[{f}_{\nu},{f}_{\lambda}]&=&J^{\mu}_{ext},
\end{eqnarray}
donde
${}^{\star}f^{\mu}(a)=\frac{1}{2}\varepsilon^{\mu\nu\lambda}F_{\nu\lambda}(a)$.
Las ecuaciones de movimiento de la teor\'{\i}a topol\'ogica masiva
corresponden al caso $\lambda=\kappa=0$. Es importante resaltar
que $J^{\mu}_{ext}$, sobre las ecuaciones de movimiento, satisface
${\cal{D}}_{\mu}J^{\mu}_{ext}=0$. Veamos
\begin{eqnarray*}
{\cal{D}}_{\mu}J^{\mu}_{ext}&=&\varepsilon^{\mu\nu\lambda}\Bigl({\cal{D}}_{\mu}{\cal{D}}_{\nu}f_{\lambda}
-\frac{\lambda}{m}[{\cal{D}}_{\mu}f_{\nu},f_{\lambda}]\Bigr)\\
&=&\Bigl[-{}^{\star}f^{\lambda}(a)-\frac{\lambda}{m}\varepsilon^{\mu\nu\lambda}{\cal{D}}_{\mu}f_{\nu},
f_{\lambda}\Bigr]
\\
&=&0,
\end{eqnarray*}
donde hemos usado la identidad
$[{\cal{D}}_{\mu},{\cal{D}}_{\nu}]f_{\lambda}=-[F_{\mu\nu}(a),f_{\lambda}]$,
la primera de las ecuaciones de (\ref{em}) y las identidades de
Jacobi para los conmutadores. Si pensaramos en algun tipo de
acoplamiento con los $f$´s la correspondiente fuente externa no se
conserva. Sin embargo el papel de $f^{\mu}$ es mas como un campo
auxiliar.

Manipulando convenientemente las ecuaciones de movimiento, cuando
no hay fuentes externas, se llega al sistema ($\lambda\neq 1$)
\begin{eqnarray}
\varepsilon^{\mu\nu\lambda}{\cal{D}}_{\nu}{f}_{\lambda}+\frac{m}{(1-\lambda)}f^{\mu}
+\frac{(\kappa-\lambda)}{2m(1-\lambda)}\varepsilon^{\mu\nu\lambda}[{f}_{\nu},{f}_{\lambda}]=0,\nonumber\\
f^{\mu}-(1-\lambda){}^{\star}f^{\mu}(a)+\frac{(\kappa-
{\lambda}^2)}{2m^2}\varepsilon^{\mu\nu\lambda}[{f}_{\nu},{f}_{\lambda}]=0.
\end{eqnarray}
En el caso que $\kappa={\lambda}^2$ tendremos que
$f^{\mu}=(1-\lambda){}^{\star}f^{\mu}(a)$, lo que nos lleva a la
ecuaci\'on de segundo orden para $a_{\mu}$
\begin{equation}
\label{ec2}
\varepsilon^{\mu\nu\lambda}{\cal{D}}_{\nu}{}^{\star}{f}_{\lambda}(a)+\overline{m}\,{}^{\star}f^{\mu}(a)
=\frac{\lambda}{2\overline{m}}\varepsilon^{\mu\nu\lambda}[{}^{\star}{f}_{\nu}(a),{}^{\star}{f}_{\lambda}(a)],
\end{equation}
donde $\overline{m}=m/(1-\lambda)$.

En (\ref{ec2}) se muestra la contribuci\'on adicional al caso de
la topol\'ogica masiva (que corresponde a $\lambda=0$) y queda
expresa la diferencia entre los dos modelos. Una ecuaci\'on igual
a esta se obtiene en el modelo no-abeliano autodual el cual fu\'e
formulado indepoendientemente por McKeon\cite{19} y por Arias, et.
al.\cite{20}.

Para el caso $\lambda=1$, al igual que en el caso abeliano,
implica que ${}^{\star}f^{\mu}(a)=f^{\mu}=0$. El caracter  masivo
de las excitaciones se hace expl\'{\i}cito si tomamos otra
derivada covariante sobre el dual de (\ref{ec2}), lo que nos lleva
a
\begin{equation}
\left(-{\cal{D}}^{\nu}{\cal{D}}_{\nu}+\overline{m}^2\right){}^{\star}f^{\mu}(a)=
\frac{(\lambda-2)}{2}\varepsilon^{\mu\nu\lambda}[{}^{\star}{f}_{\nu}(a),{}^{\star}{f}_{\lambda}(a)]+
\frac{\lambda}{\overline{m}}[{\cal{D}}_{\nu}{}^{\star}{f}^{\mu}(a),{}^{\star}{f}^{\nu}(a)],
\end{equation}
corroborando que la masa de las excitaciones es $\vert
m/(1-\lambda)\vert$ como en el caso abeliano.

Hemos, entonces mostrado como se conectan los modelos vectoriales
propuestos por Hagen por la v\'{\i}a del mecanismo de
autointeracci\'on. Tambien qued\'o explicita la no equivalencia
entre los modelos no-abelianos de Hagen y de la teor\'{\i}a
topol\'ogica masiva para un caso particular. En el caso que
$\lambda=1$ ($\kappa=1$) la teor\'{\i}a de Hagen s\'olo describe
estados globales ($F_{\mu\nu}(a)=0$) los cuales son sensibles a la
topolog\'{\i}a del espacio base.

Este trabajo est\'a enmarcado dentro del Proyecto de Grupo
G-2001000712 del FONACIT.


\end{document}